\documentclass[3p,times,twocolumn]{elsarticle}
 \biboptions{comma,sort&compress}
 
\usepackage{graphicx}
\usepackage{braket}
\usepackage{amsmath,stackrel}
\usepackage{here}
\usepackage{ecrc}

\volume{00}

\firstpage{1}

\journalname{Nuclear and Particle Physics Proceedings}

\runauth{}

\jid{nppp}

\jnltitlelogo{Nuclear and Particle Physics Proceedings}

\usepackage{amssymb}

\usepackage[figuresright]{rotating}

\newcommand{\MeV}{\,{\rm MeV}}
\newcommand{\GeV}{\,{\rm GeV}}

\newcommand{\QCD}{{\textrm{\scriptsize QCD}}}

\newcommand{\BG}{{\textrm{\scriptsize BG}}}
\newcommand{\TF}{{\textrm{\scriptsize TF}}}
\newcommand{\q}{{\tilde{q}}}

\begin{document}

\begin{frontmatter}

\title{ Tsallis statistics, fractals and QCD $^*$}
 \cortext[cor0]{Talk given at 23th International Conference in Quantum Chromodynamics (QCD 20,  35th anniversary),  27 - 30 october 2020, Montpellier - FR}
 \author[label1]{Airton Deppman}
\ead{deppman@usp.br}
\address[label1]{Instituto de F\'{\i}sica, Universidade de S\~{a}o Paulo, Rua do Mat\~ao 1371-Butant\~a, S\~ao Paulo-SP, CEP 05580-090, Brazil}
 \author[label2]{Eugenio Meg\'{\i}as\fnref{fn1}}
   \fntext[fn1]{Speaker, Corresponding author.}
\ead{emegias@ugr.es}
\address[label2]{Departamento de F\'{\i}sica At\'omica, Molecular y Nuclear and Instituto Carlos I de F\'{\i}sica Te\'orica y Computacional, Universidad de Granada, Avenida de Fuente Nueva s/n, 18071 Granada, Spain}
 \author[label3]{D\'ebora P. Menezes}
\ead{debora.p.m@ufsc.br}
\address[label3]{Departamento de F\'{\i}sica, CFM-Universidade Federal de Santa Catarina, Florian\'opolis, SC-CP. 476-CEP 88.040-900, Brazil}

\pagestyle{myheadings}
\markright{ }
\begin{abstract}
We study the non-extensive Tsallis statistics and its applications to QCD and high energy physics, and analyze the possible connections of this statistics with a fractal structure of hadrons. Then, we describe how scaling properties of Yang-Mills theories allow the appearance of self-similar structures in gauge fields, which actually behave as fractals. The Tsallis entropic index, $q$, is deduced in terms of the field theory parameters, resulting in a good agreement with the value obtained experimentally.

\end{abstract}

\end{frontmatter}

\section{Introduction}
\label{sec:introduction}

The description of Quantum Chromodynamics (QCD) in the hot and dense regimes has experienced important advances in recent years. The study of hadronic systems in a medium has motivated the introduction of several approaches, including lattice studies~\cite{Borsanyi:2010cj}, chiral quark models~\cite{Megias:2004hj}, and hadron resonance gas (HRG) models~\cite{Huovinen:2009yb,Megias:2012kb}. The latter are based on the findings of R. Hagedorn, who proposed a self-consistent thermodynamical approach to QCD formulated in terms of Boltzmann-Gibbs (BG) statistics~\cite{Hagedorn:1965st,Hagedorn:1984hz}, allowing a description of the confined phase as a multi-component gas of non-interacting massive stable particles. In spite of its remarkable success to describe the equation of state of QCD, when confronted to $pp$ experimental data, the Hagedorn's approach led to some discrepancies that motivated a modification in the formalism. Tsallis statistics was introduced as a generalization of BG statistics by considering a non-additive form of entropy, and it has found wide applicability in the last few years apart from high energy physics, see e.g.~\cite{Tempesta:2011vc,Kalogeropoulos:2014mka}.

On the other hand, fractals are complex systems with internal structure presenting self-similarity~\cite{Mandelbrot}. One important aspect of fractals is the fractal dimension, which describes how the measurable aspects of the system change with scale. A direct consequence is the power-law behavior of distributions observed for~fractals.  

The goal of this manuscript is to show the link between Tsallis statistics, fractals and renormalization group (RG) invariance of Yang-Mills field (YMF) theories, with applications to the phenomenology of QCD.

\section{Tsallis statistics in high energy physics}
\label{sec:Tsallis_HEP}

\begin{figure}[t] 
\begin{center}
\includegraphics[width=0.35\textwidth]{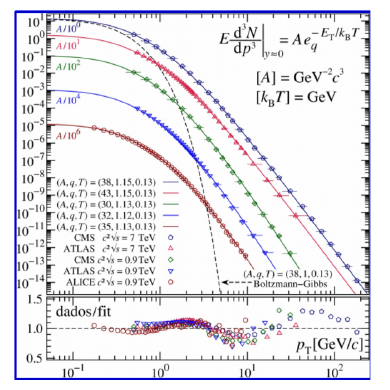} 
\end{center}
\vspace{-0.4cm}
\caption{\it Fitting of the experimental data for $p_\perp$ distribution of the abundance of different hadron species, by considering the Tsallis factor of Eq.~(\ref{eq:PTsallis}). The NESCT fits well data over $15$ orders of magnitude.
}
\label{fig:pT}
\end{figure}

When the Hagedorn's approach was applied to $pp$ collisions, it predicted the transverse momentum distribution of the particle production of hadrons, given by
\begin{equation}
\frac{d^2{\mathcal N}}{dp_\perp dy} = g V \frac{p_\perp m_\perp}{(2\pi)^2} e^{-m_\perp/(k_B T)} \,, \label{eq:dN_Hagedorn}
\end{equation}
where $g$ is a constant, $V$ is the volume of the system, $m_\perp = (p_\perp^2 + m^2)^{1/2}$, $y$ is the rapidity, and $k_B$ is the Boltzmann constant. However, this exponential distribution in energy and momentum turned out to be in disagreement with experimental data, as these tend to behave instead as a power-law, cf. Fig.~\ref{fig:pT}. The non-extensive self-consistent thermodynamics (NESCT) was proposed as an extension of the Hagedorn's theory to non-extensive statistics~\cite{Deppman:2012us}, and it was based on the Tsallis factor
\begin{equation}
P(\varepsilon) = A \left[ 1+(q-1)\frac{\varepsilon}{k_B T} \right]^{-\frac{1}{q-1}} \,, \label{eq:PTsallis}
\end{equation}
instead of the exponential BG factor, $P_{\BG}(\varepsilon) = e^{-\varepsilon/(k_B T) }$.
This extended theory allowed to reproduce the distribution of all the species produced in $pp$ collisions with high accuracy, cf. Fig.~\ref{fig:pT}, leading to $q = 1.14 \pm 0.01$ and $T = 62 \pm 5 \, \MeV$~\cite{Marques:2012px,Marques:2015mwa}. Moreover, the NESCT predicts a power-law behavior for the hadron spectrum~$\rho(m)= \rho_o \cdot \left[1+(q-1) \frac{m}{T} \right]^{1/(q-1)}$~\cite{Deppman:2012us}. A numerical comparison with the hadron spectrum from the PDG~\cite{Zyla:2020zbs}, leads to an important improvement with respect to the distribution $\rho(m) = \rho_0 \cdot e^{m/T_H}$ proposed by Hagedorn, specially at the lowest masses, cf. Ref.~\cite{Marques:2012px}.

\section{Tsallis statistics and QCD thermodynamics}
\label{sec:definition}

\begin{figure}[t] 
\begin{center}
\includegraphics[width=0.43\textwidth]{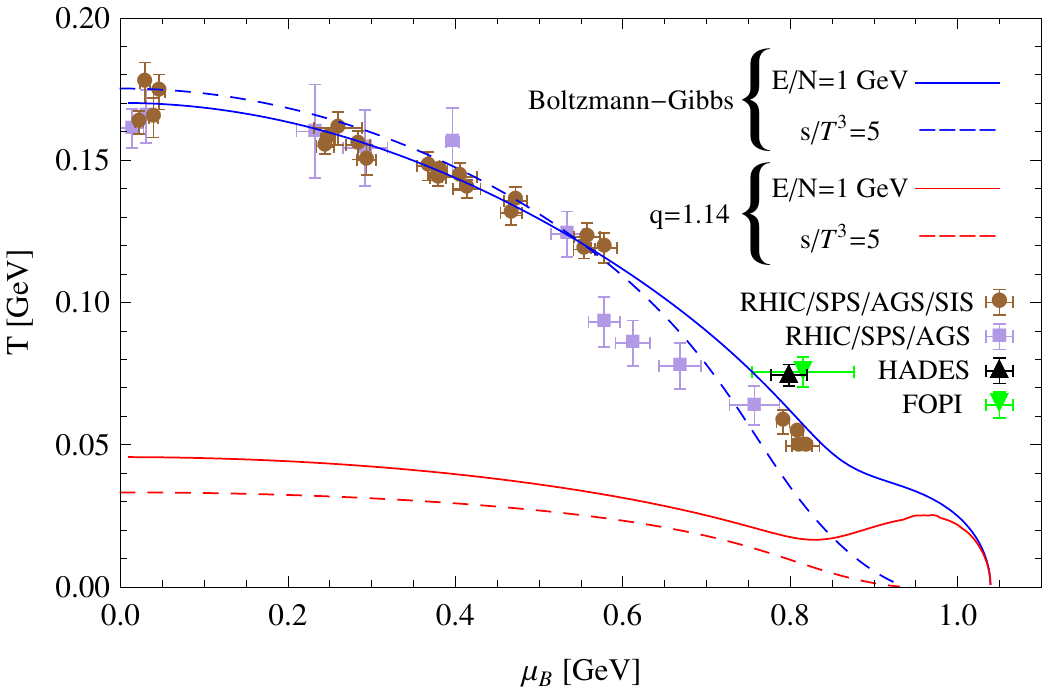} 
\end{center}
\vspace{-0.4cm}
\caption{\it Chemical~freeze-out~line~$T = T(\mu_B)$. We plot the result by using BG statistics, and~Tsallis statistics with $q=1.14$. 
}
\label{fig:thermodynamics}
\end{figure}

Tsallis statistics is a generalization of BG statistics, where the
entropy is given by~\cite{Tsallis:1987eu}
\begin{equation}
S_q = -k_B \log_q^{(-)} p(x) \,, \label{eq:Sq}
\end{equation} 
where $p(x)$ is the probability of $x$ to be observed, and $q$ is the entropic index. This leads to a non-extensive thermodynamics that is described in terms of the $q$-exponential, $e_q^{(\pm)}(x)=[1 \pm (q-1)x]^{\pm1/(q-1)}$, and  $q$-logarithmic, $\log^{(\pm)}_q(x)=\pm (x^{\pm(q-1)}-1)/(q-1)$, functions. A direct consequence of Eq.~(\ref{eq:Sq}) is the non-additivity of entropy, since for two independent systems, $A$ and $B$, the entropy of the combined system is 
\begin{equation}
S_{A+B} = S_A + S_B + k_B^{-1} (q-1)S_A S_B \,,
\end{equation}
where $q$ measures the degree of non-additivity. As $q \to 1$ the BG statistics is recovered. This formalism has been applied to QCD at finite temperature and chemical potential. The grand-canonical partition function for a non-extensive ideal quantum gas is given by~\cite{Megias:2014tha,Megias:2015fra}
\begin{eqnarray}
\hspace{-1.2cm} &&\log Z_q(V,T,\mu)   \nonumber \\
\hspace{-1.2cm} &&\quad = -\xi V\int \frac{d^3p}{{(2\pi)^3}} \sum_{r=\pm}\Theta(r x)\log^{(-r)}_q\left(\frac{ e_q^{(r)}(x)-\xi}{ e_q^{(r)}(x)}\right)  \,, \label{eq:Zq}
\end{eqnarray}
where $x = (E_p - \mu)/(k_B T)$, $E_p$ is the particle energy, $\mu$ is the chemical potential, $\xi = \pm 1$ for bosons(fermions), and $\Theta$ is the step function. Tsallis statistics has been used to successfully describe the thermodynamics of QCD in the confined phase by using the HRG approach~\cite{Hagedorn:1984hz}. In this approach, the partition function~is
\begin{equation}
\log Z_q(V,T,\{\mu_{Q}\})=\sum_{i \in \textrm{hadrons}} \log Z_q(V,T,\mu_{Q_i})\,, \label{eq:logZ}
\end{equation}
where $\mu_{Q_i}$ refers to the chemical potential of charge $Q$ for
the $i$-th hadron. The states are usually taken as the conventional
hadrons listed in the review by the Particle Data Group
(PDG)~\cite{Zyla:2020zbs}. Using the arguments of
Refs.~\cite{Cleymans:1999st,Tawfik:2005qn}, the phase transition line
in the $T - \mu_B$ diagram, with $\mu_B$ the baryonic chemical
potential, can be determined by the conditions $\langle E\rangle /
\langle N \rangle \simeq 1 \GeV$ or $s/T^3 \simeq 5$. The results,
displayed in Fig.~\ref{fig:thermodynamics}, show that the freeze-out
line spams over the region of $0<\mu_B<1039.2\,\MeV$, with an
inflection for $\mu_B \sim 0.9 \, \GeV$ related to the sharp increase
in the baryon density as $\mu_B$ approaches the proton/neutron
mass~$m_{p,n} \simeq 0.94 \, \GeV$.

\section{Thermofractals}
\label{sec:thermofractals}

The emergence of the non-extensive behavior has been attributed to
different causes: i) long-range interactions, correlations and memory
effects~\cite{Borland:1998}; ii) temperature fluctuations; and iii)
finite size of the system. We will study in this section a natural
derivation of non-extensive statistics in terms of thermofractals.

\subsection{Fractals and self-similarity}
\label{subsec:fractals}

Fractals are defined in terms of their self-similar properties at different scales. A scaling transformation changes the size of a physical system by a scaling factor, $\lambda$, leading this to some specific behaviors with $\lambda$ of the physical quantities characterizing the system. On the other hand, a self-similar system is a system which is similar to a part of itself. If length is reduced by the scaling factor as $\ell(\lambda) = \frac{\ell_0}{\lambda}$, then a system can be filled by $N$ smaller self-similar systems, where~$N(\lambda) = N_0 \lambda^D$. The fractal dimension, $D$, is then defined as
\begin{equation} 
D = \lim_{\lambda\to \infty} \frac{\log N(\lambda)}{\log \lambda} \,.
\end{equation}
A classical example of fractal is the length of coastlines, as it
depends on the resolution, i.e.~$L(\lambda) = N(\lambda) \ell(\lambda)
= L_0 \cdot \lambda^{D-1}$, where~$L_0 = N_0 \ell_0$ is the measured
length at some initial resolution. One can see that the increase of
$L$ with $\lambda$ is indicative of a fractal dimension $D > 1$.

\subsection{Thermofractals}
\label{subsec:thermofractals}

These are systems in thermodynamical equilibrium presenting the following properties~\cite{Deppman:2016fxs}:
\begin{enumerate}[(i)]
\item Total energy is given by~$U = F + E$, where $F$ is the kinetic energy, and $E$ is the internal energy of $N$ constituent subsystems, so that $E = \sum_{i=1}^N \varepsilon_i^{(1)}$.
\item The constituent particles are thermofractals. This means that the energy distribution $P_{\tiny\mathrm{\tiny TF}\tiny}(E)$ is self-similar, i.e. at level $n$ of the hierarchy of subsystems, $P_{\TF (n)}(\varepsilon)$ is equal to the distribution in any other level, $P_{\TF (n)}(\varepsilon) \propto P_{\TF (n+m)}(\varepsilon)$.
\end{enumerate}
In thermofractals, the phase space must include momentum d.o.f. $(\propto f(F))$ of free particles as well as internal d.o.f.~$(\propto f(\varepsilon))$. According to property (ii) of self-similar thermofractals, the energy distribution is given by
\begin{equation}
\hspace{-0.6cm} P_{\TF (0)}(U) dU = A^\prime  F^{\frac{3N}{2}-1} e^{-\alpha F/(k_BT)} dF  \left[ P_{\TF (1)}(\varepsilon) \right]^\nu d\varepsilon 
\end{equation} 
with $\alpha = 1 + \frac{\varepsilon}{(k_BT)}$ and $ \frac{\varepsilon}{k_BT} = \frac{E}{F} $, and $\nu$ should be determined. By imposing~$P_{\TF (0)}(U) \propto P_{\TF (1)}(\varepsilon)$, one finds
\begin{equation}
P_{\TF (n)}(\varepsilon) = A_{(n)} \cdot \bigg[1+ (q-1) \frac{\varepsilon}{k_B\tau}\bigg]^{-\frac{1}{q-1}}  \,. \label{eq:selfsimilar}
\end{equation}
The distribution of thermofractals then obeys Tsallis statistics with $q-1 = 2(1-\nu)/(3N)$ and $\tau = N (q-1) T$.

\section{Fractal structures in Yang-Mills fields}
\label{sec:YMT}

We have discussed in previous sections that QCD can be described by
Tsallis statistics, and that thermofractals obey this statistics. We
will address now whether it is possible a thermofractal description of
YMF theory.

\subsection{Renormalization of gauge fields}
\label{subsec:renormalization}

\begin{figure}[t]
\begin{center}
\includegraphics[width=0.27\textwidth]{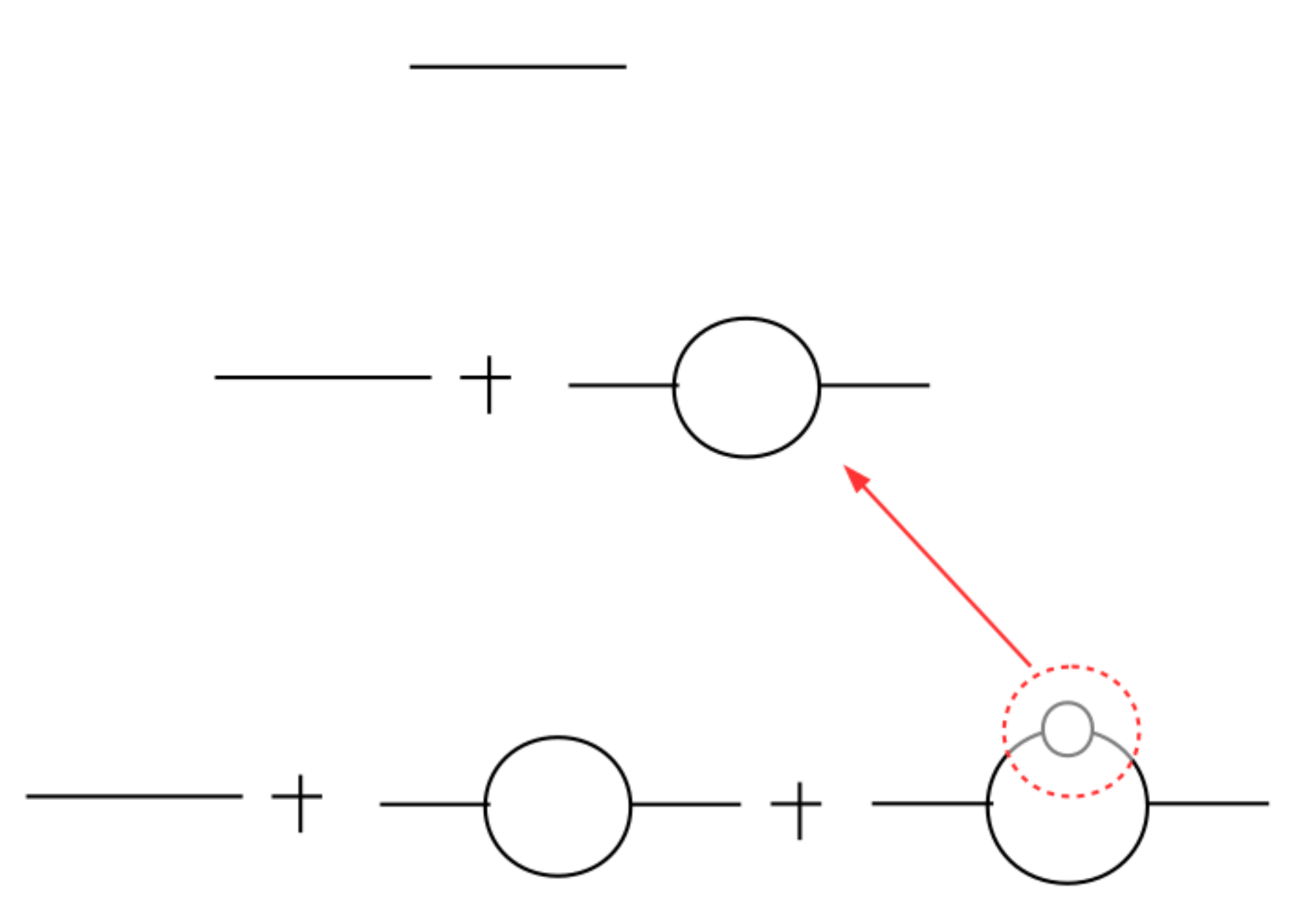} \hspace{0.5cm}
\includegraphics[width=0.15\textwidth]{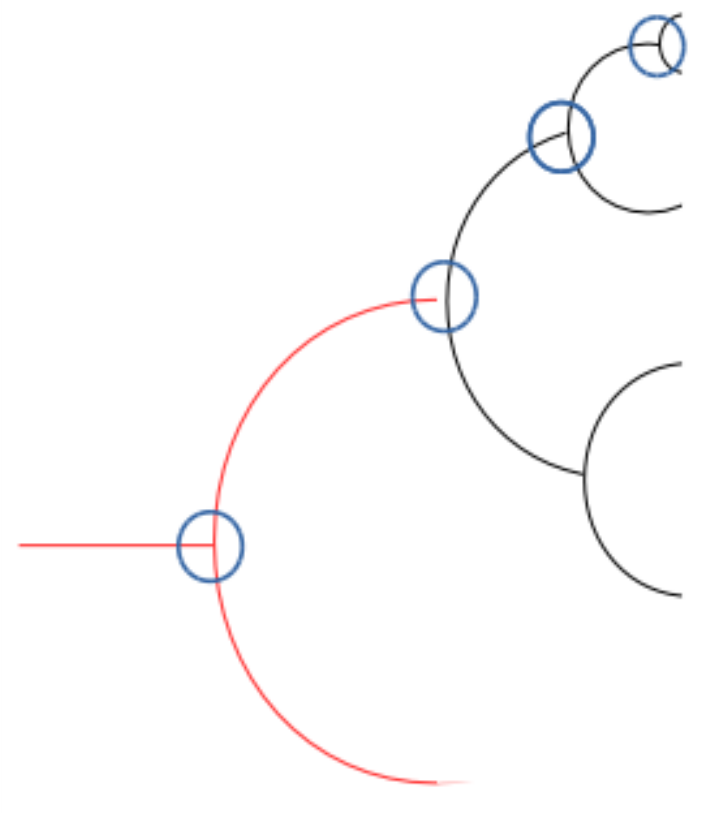}
\end{center}
\vspace{-0.4cm}
 \caption{\it Left panel: Diagrams showing the scaling properties of
   YMF. It is shown the loops at different orders in the perturbative
   expansion. Right panel: Diagrams of multiparticle production in
   $pp$ collisions. The initial parton (first black line from the
   left) may be considered as a constituent of another parton (first
   red line from the left).  The circles indicate the effective vertex
   couplings, given by Eq.~(\ref{eq:g_effective}).
}
\label{fig:scaling_multiparticle}
\end{figure}

The YMF theory was shown to be renormalizable in Ref.~\cite{tHooft:1972tcz}. This means that vertex functions, that are regularized to avoid the ultraviolet divergences, are related to the renormalized vertex functions, $\Gamma(p,m,g)=\lambda^{-D} \Gamma(p,\bar{m},\bar{g})$, where $\bar{m}$ and $\bar{g}$ are renormalized parameters, and $\lambda$ is a scale transformation parameter, i.e.,~$p^\mu \to p^{\prime\, \mu} = \lambda p^\mu$~\cite{Dyson:1949ha,GellMann:1954fq}. This property is described by the RG equation, also known as Callan-Symanzik (CS) equation~\cite{Callan:1970yg,Symanzik:1970rt}
\begin{equation}
\left[M\frac{\partial}{\partial M}  + \beta_{\bar g} \frac{\partial}{\partial \bar{g}}  + \bar \gamma \right]\Gamma=0  \,,
\end{equation}
where $M$ is the scale parameter of the theory, and the beta function is defined as $\beta_{\bar g} = M \frac{\partial \bar{g}}{\partial M}$. As it is shown in Fig.~\ref{fig:scaling_multiparticle} (left), RG invariance means that, after proper scaling, the loop in a higher order graph in the perturbative expansion is identical to a loop in lower orders. This is a direct consequence of the CS equation, and it is indicative of self-similar properties of gauge fields.

A physical situation in which these self-similar properties emerge is in multiparticle production in $pp$ collisions~\cite{Konishi:1978ks}. We show in Fig.~\ref{fig:scaling_multiparticle} (right) a typical diagram in which two partons are created in each vertex. The line of the respective field represents an effective particle, and the vertices are related to the creation of an effective parton. Another example is the hadron structure: as in multiparticle production, too many complex graphs should be considered when studying the hadron~at a high resolution~\cite{Konishi:1978ks}. Then, calculations are limited either to the first leading orders, or to lattice QCD.

\subsection{Fractal structure of gauge fields}
\label{subsec:fractal_gauge_fields}

We will consider that at any scale the system can be described by an ideal gas of particles with different masses, i.e. masses might change with the scale. Let us study the time evolution of an initial partonic state
\begin{equation}
\ket{\Psi(t)}=e^{-iHt} \ket{\Psi_o} \,.
\end{equation}
We will introduce two different bases: i) $\ket{\Psi_n}$ represent states with $n$ interactions, and ii) $\ket{\psi_N}$ are states with $N$~partons. The number of particles in the state $\ket{\Psi_n}$ is not directly related to~$n$, since there might by annihilation of particles at any order. We will denote by $\tilde{N}$ the number of particles created or annihilated at each interaction.
Let us study the probability to find a state with one parton with mass between $m_o$ and $m_o + dm_o$, and momentum between $p_o$ and $p_o + dp_o$. This is given by
\begin{eqnarray}
\hspace{-1.2cm} && {\tilde P}(\varepsilon) d^4p_0 dE \equiv \braket{\gamma_o,m_o,p_o, \dots|\Psi(t)} \nonumber \\
\hspace{-1.2cm} &&\;= \sum_n  \sum_{N} \braket{\Psi_n|\Psi(t)} \braket{\psi_N|\Psi_n} \braket{\gamma_o,m_o,p_o,\dots|\psi_N} \,.
\end{eqnarray}
The factor~$\braket{\Psi_n|\Psi(t)}=G^n P(E) dE$ is related to the probability that an effective parton with energy between $E$ and $E+dE$ and probability distribution $P(E)$, will evolve in such a way that at time $t$ it will generate an arbitrary number of secondary effective partons in a process with $n$ interactions. The bracket~$\braket{\psi_N|\Psi_n} \stackrel[n \gg 1]{\simeq}{}   \left( \frac{N}{n(\tilde{N}-1)} \right)^4$ is the probability to get the configuration with $N$ particles after $n$ interactions. Finally, the last bracket can be calculated statistically, leading to the following result~\cite{Deppman:2019yno}
\begin{equation}
\braket{\gamma_o,m_o,p_o,\dots|\psi_N} = A(N) P_N\left( \frac{\varepsilon_j}{E} \right) d^4\left( \frac{p_j}{E} \right) \,,
\end{equation}
with~$A(N) = \frac{1}{8\pi}\frac{\Gamma(4N)}{\Gamma(4(N-1))}$ and $P_N(x) = (1-x)^{4N-5}$, while $p^\mu_j$ is the four-momentum of particle $j$ inside the system of $N$ particles, and $\varepsilon_j = p_j^0$. Let us consider that the system with energy $E$ in which the parton with energy~$\varepsilon_j$ is one among $N$ constituents, is itself a parton inside a larger system with energy $\cal{M}$. Then self-similarity implies that~$\tilde{P}\left( \frac{\varepsilon_j}{E} \right) \propto P\left( \frac{E}{\cal{M}} \right)$, and it can be shown that 
\begin{equation}
P\left( \frac{\varepsilon}{\lambda}\right) = \left[ 1 + (q-1) \frac{\varepsilon}{\lambda} \right]^{-\frac{1}{q-1}} \,, \label{eq:P}
\end{equation}
where $q-1 = (1-\nu)/(4N-5)$, and $\lambda = (q-1)\Lambda$ is a reduced scale. This power-law distribution corresponds to the one derived in Eq.~(\ref{eq:selfsimilar}) for thermofractals, cf.~\cite{Deppman:2016fxs,Deppman:2017fkq}. Notice that Eq.~(\ref{eq:P}) describes how the energy received by the initial parton flows to its internal d.o.f.. This suggests that at each vertex, this distribution plays the role of an effective coupling (cf. Fig.~\ref{fig:scaling_multiparticle} (right))
\begin{equation}
\bar{g}= \prod_{i=1}^{\tilde{N}} G \left[1 + (q-1)\frac{\varepsilon_i}{\lambda}\right]^{-\frac{1}{q-1}} \,. \label{eq:g_effective}
\end{equation}

\subsection{Beta function: determination of $q$}
\label{eq:determination_q}

\begin{figure}[t] 
\begin{center}
\includegraphics[width=0.23\textwidth]{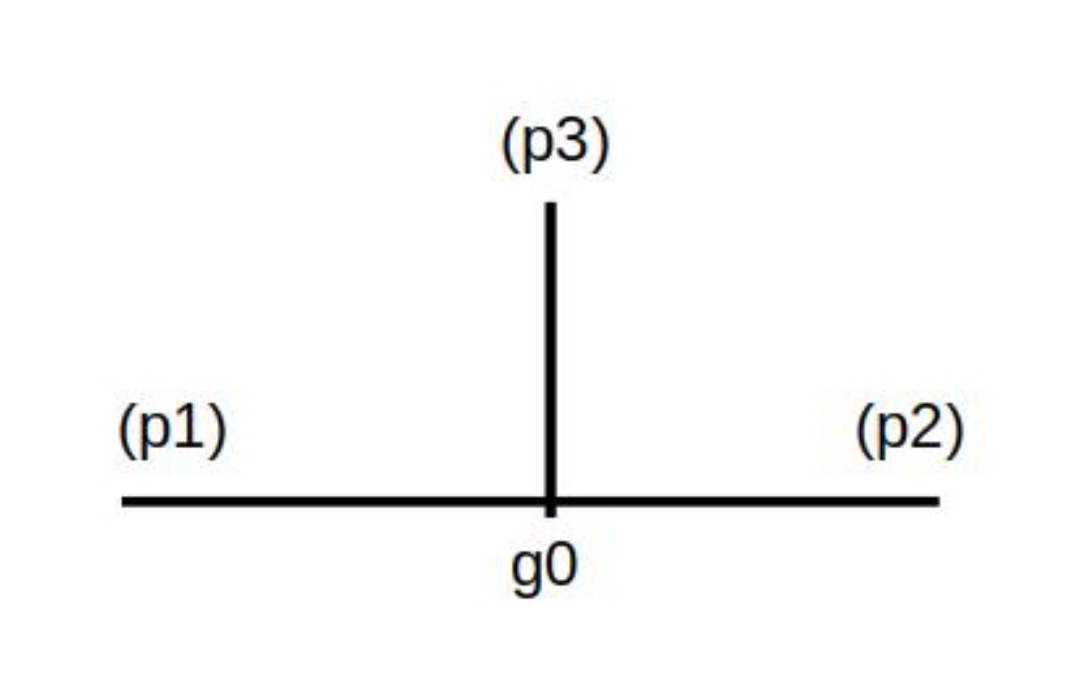} 
\includegraphics[width=0.23\textwidth]{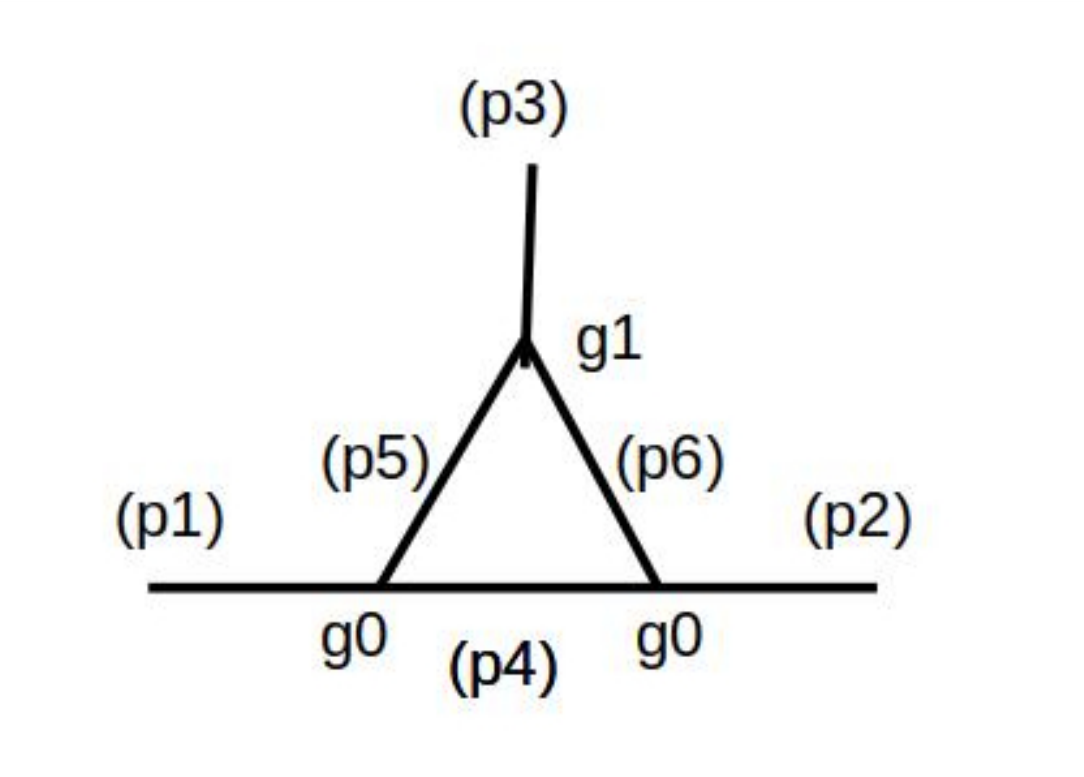}
\end{center}
\vspace{-0.4cm}
\caption{\it Vertex functions at scale $\lambda_o$ (left) and $\lambda$ (right).
}
\label{fig:vertex_functions}
\end{figure}

The renormalized vertex functions together with the CS equation were used to derive the beta function of QCD, leading to the 1-loop result~\cite{Politzer:1974fr,Gross:1974cs}
\begin{equation} 
\beta_{\QCD} = - \frac{1}{16\pi^2} \left[ \frac{11}{3}c_1 - \frac{4}{3} c_2 \right] \bar{g}^{3} \,, \label{eq:beta_QCD}
\end{equation}
where $c_1 = N_c$ and $c_2 = N_f/2$. The beta function derived within our formalism can be computed as follows. Let us consider a vertex in two different orders, cf. Fig.~\ref{fig:vertex_functions}. From a comparison of the vertex function at scale $\lambda_o$, i.e. $\Gamma_o=\braket{\gamma_2 p_2 , \gamma_3 p_3|\bar{g}(\lambda_o)\, e^{-iH_ot}| \gamma_1 p_1}$, with the vertex function at scale $\lambda$ which contains one additional loop, $\Gamma$, one can identify the effective coupling $\bar{g}$. By using the result of Eq.~(\ref{eq:g_effective}) with~$\lambda = \lambda_o/\mu$, where $\mu$ is a scaling factor, together with previous considerations, one can calculate the 1-loop beta function, leading to
\begin{equation}
\beta_{\bar{g}} = \mu \frac{\partial \bar{g}}{\partial \mu}= - \frac{1}{16\pi^2} \frac{1}{q-1} g^{\tilde{N}+1} \,,
\end{equation}
with $\tilde{N} = 2$ in YMF theory. By comparing with the QCD result of Eq.~(\ref{eq:beta_QCD}), one obtains $q=1.14$ when using $N_c = N_f/2 = 3$, in excellent agreement with the experimental data analyses discussed in Secs.~\ref{sec:Tsallis_HEP} and \ref{sec:definition}.  

\section{Applications}
\label{sec:applications}

The main problem is to verify to what extent the fractal structure can
describe experimental data in high energy physics. One of the most
used techniques to unveil fractal dimensions from $p_\perp$
distributions~\cite{Cleymans:2011in,Wong:2015mba,Marques:2015mwa} is
the intermittency analysis. Let $\mathcal N(y)$ be the rapidity
distribution, let $k$ denote the number of particles detected in a
narrow window $\delta y$ in one event, and let $Q_k$ be the number of
times that the multiplicity $k$ recurs in $\mathcal N$ events. One can
define the moments of order $\q$ by
\begin{equation}
G_{\q} = \frac{\sum_{k=0}^\infty k^\q Q_k}{\left(\sum_{k=0}^\infty k Q_k\right)^\q} \propto \delta y^{(\q-1)D} \,,
\end{equation}
where $D$ is the Hausdorff fractal dimension. The analysis of these moments leads to $D = 0.65$~\cite{Azhinenko:1989jr,Albajar:1992hr,Sarkisian:1993mf,Rasool,Singh:1994uh,Ghosh:1998gw}. From the thermofractal structure it can be obtained $\mathcal N \propto E^{D-1}$, leading to $D = 0.69$ when $q = 1.14$ is considered, in agreement with the intermittency analysis, cf. Ref.~\cite{Deppman:2016fxs}.

The application of the NESCT to systems with finite chemical potential has allowed the application of the thermodynamics properties of hadrons to the study of neutron stars~\cite{Lavagno:2011zm,Menezes:2014wqa}. It is found that the internal temperature of the stars decreases with the increase of~$q$. This aspect may have important consequences when the star evolves from a hot and lepton rich object to a cold and deleptonized compact star. Finally, the hadron structure can benefit from the thermofractal theory presented here. Some models have already used Tsallis statistics to introduce the fractal aspects of QCD in the description of hadron structure~\cite{Cardoso:2017pbu,Andrade:2019dgy}.

\section{Conclusions}
\label{sec:Conclusions}

We have investigated the structure of a thermodynamical system presenting fractal properties showing that it naturally leads to Tsallis non-extensive statistics. By using a formalism that reflects the self-similar features of fractals, we have shown that renormalizable field theories can lead to fractal structures. Within this picture, we have computed the effective coupling and the beta function of QCD, giving a prediction for the entropic index, $q$, which is in good agreement with the value obtained by fitting Tsallis distributions to experimental data. Beyond the phenomenological success of this description, let us remark that self-similarity in gauge fields leads to interesting properties, as e.g. fractal structure, recursive calculations at any order, reconciliation of Hagedorn's theory with QCD, and fractal dimension in multiparticle production (see Ref.~\cite{Deppman:2020jzl} for a recent review). The study of all these features deserves further investigation.

\section*{Acknowledgements} 
The research of A.D. and D.P.M. were funded by the Conselho Nacional
de Desenvolvimento Cient\'{\i}fico e Tecnol\'ogico (CNPq-Brazil) and
by Project INCT-FNA Proc. No. 464898/2014-5, and by FAPESP under grant
2016/17612-7. The work of E.M. is funded by the Spanish MINEICO under
Grant FIS2017-85053-C2-1-P, by the FEDER Andaluc\'{\i}a 2014-2020
Operational Programme under Grant A-FQM-178-UGR18, by Junta de
Andaluc\'{\i}a under Grant FQM-225, and by the Consejer\'{\i}a de
Conocimiento, Investigaci\'on y Universidad of the Junta de
Andaluc\'{\i}a and ERDF under Grant SOMM17/6105/UGR. The research of
E.M. is also supported by the Ram\'on y Cajal Program of the Spanish
MINEICO under Grant RYC-2016-20678.

\end{document}